\documentclass[12pt]{article}
\usepackage[dvips]{epsfig}
\begin{document}
\thispagestyle{empty}

\LARGE
\begin{center}
{\bf Is binary sequential decay compatible
with the fragmentation of nuclei at high energy ?}

\normalsize
\bigskip
{\sf P.~Wagner$^{*}$\footnote{e-mail: wagner@sbgvsg.in2p3.fr},
J.~Richert$^{**}$\footnote{e-mail: richert@lpt1.u-strasbg.fr},
V. A. Karnaukhov$^{***}$\footnote{e-mail: karna@nusun.jinr.dubna.su}
and H.~Oeschler$^{****}$\footnote{e-mail: h.oeschler@gsi.de}}

\bigskip
$^{*}$ Institut de Recherches Subatomiques, BP28, \\
67037 Strasbourg Cedex 2, France

$^{**}$ Laboratoire de Physique Th\'eorique,
3, rue de l'Universit\'e,\\ 67084 Strasbourg Cedex, France

$^{***}$ Joint Institute for Nuclear Research,\\ 141980 Dubna, Russia

$^{****}$ Institut f\"ur Kernphysik, Technische Universit\"at
Darmstadt,\\ 64289 Darmstadt, Germany

\end{center}

\vspace{1cm}

\begin{abstract}
We use a binary sequential decay model in order to describe the
fragmentation of a nucleus induced by the high energy collisions of protons
with Au nuclei. Overall agreement between measured and calculated
physical observables is obtained. We evaluate and analyse the decay times
obtained with two different parametrisations of the decay rates and
discuss the applicability of the model to high energy fragmentation.
\end{abstract}

\normalsize
{\sloppy

Intermediate mass fragment yields. Reaction time.

\bigskip
The time evolution of a nuclear system which undergoes
fragmentation is still an open question. It is, of course, strongly
correlated to the dynamical evolution of the process. Hence, 
time arguments may help to differentiate between various mechanisms.
Experimental quantities sensitive to the evolution are of
great interest. The aim of the present study is an attempt to test a
simple model describing sequential decay by working out
specific observables which can be confronted with available
experimental results. Characteristic times given
by the model are compared to experimental and theoretical estimates.
This work complements a recent analogous study in which simultaneous
decay has been investigated~\cite{Oe_Botvina}. 

\bigskip
The experiment concerns the fragmentation produced by the
impinging of a proton on a Au target at 8.1 GeV incident
energy. 
After the fast cascade process a system with mass A $\simeq$ 160
is left over. This sytem decays into fragments and has been
investigated experimentally~\cite{Avdeyev}.

\bigskip
Among many different fragmentation mechanisms the disassembly through 
sequential binary decay has been worked out in different forms
~\cite{Friedman,Durand,Pal}. Here we introduce a
time-dependent description which has been proposed some time
ago~\cite{Richert}. We start with an excited system (here
A$\simeq$ 160) which decays into smaller species
by means of a binary sequential decay process. The binary
decay is governed by transition rates, either taken from the Weisskopf
(WTR) detailed-balance principle~\cite{Weisskopf} or from the 
transition-state theory (STR) proposed by Swiatecki~\cite{Swiatecki}.
The process is numerically simulated as discussed in~\cite{Richert}. 
At each time step,
any composite fragment which has been formed can decay into
two smaller species corresponding to a 
decay channel determined by means of a random procedure.
The fragments fly apart in a randomly chosen direction due
to the Coulomb repulsion which acts between them, in such
a way that they never overlap in space. The process
stops when all the generated species can no longer disassemble
because their energy lies below the lowest decay threshold.
The difference between the present calculation and the \mbox{original}
one~\cite{Richert} consists of the introduction of the Coulomb
interaction which acts between all fragments at any time. As
time flows these fragments fly apart along classical trajectories.
The simulation is repeated in order to get significant statistics.

\bigskip
We have applied the model to the calculation of different observables
by choosing an initial excitation energy $E/A = 5~MeV$ which lies
in the range of the expected experimental excitation energies.
Calculations have been made for $E/A = 4~MeV$ and the results are not 
significantly different. In the sequel we concentrate on the case where
$E/A = 5~MeV$ for which the intermediate mass fragment (IMF, i.e.
$3 \leq Z \leq 20$)
multiplicity is closest to the experimental value.
Fig.~1a shows the charge distribution obtained with the two types of
transition rates. Both calculations show qualitative agreement with
the experiment. Fig.~1b concerns the energy distribution of C isotopes.
For energies of the carbons lower than 60 MeV, the calculated yields
follow nicely the experimental ones. Discrepancy with the experiment 
appears for higher carbon energies. The measured high energy tail of
the distribution cannot be reproduced by the models. We shall come back
to this point in the sequel. A second observable is shown in Figs.~2,
where we compare the energy distribution of C isotopes for 
different multiplicities  $M_{\mathrm{IMF}}$ of intermediate
mass fragments. The evolution of the calculated spectra with the
multiplicity looks similar to the experimental ones.
One has to keep in mind that $M_{\mathrm{IMF}}$ in the lower
part is the total $M_{\mathrm{IMF}}$ while in the upper part
$M_{\mathrm{A}}$ is the number of measured IMF's which is of course
smaller than $M_{\mathrm{IMF}}$ due to detector efficiency.
In particular, the model reproduces the shift to the left of the
maxima of the curves with increasing IMF multiplicity.
This shift can be qualitatively understood as being due to the
fact that the energy which is available for the emitted carbon
diminishes with the generation of an increasing number of
IMF's which take up part of the total conserved energy.
More details about these distributions can be seen in Figs.~3.
Comparing Figs.~3a and 3b one observes that the energies
of the maxima agree quite nicely. As for the mean energies, the 
experimental values are rather constant while the calculated ones 
decrease substantially with $M_{\mathrm{IMF}}$ when one uses WTR rates.
This may arise from the fact that in the experiment various
excitation energies are involved and hence, in the experimental
curves, both the maxima and the slopes are changing. The
results obtained with the MMMC~\cite{Gross} and the SMM~\cite{Bondorf}
models are shown in Figs.~3c and 3d respectively.
As it can be seen MMMC simulations lead to rather
constant mean energies, although their absolute values
are somewhat smaller than the experimental ones. The SMM simulations
show to some extent the decreasing trend of the sequential decay model.
If one knows that the SMM allows for the existence of a final
stage which does not appear in MMMC calculations and in which the
system evaporates particles from the existing excited fragments,
it may be tempting to believe that the trend observed in these mean
energies could be related to the sequential character of the
disassembly. The calculations with STR which are not shown
behave similarly, the mean values being closer to those shown in Fig.~3c.

\bigskip
From the present results, one would conclude that the present model
works quite satisfactorily. This has also been
observed in the application of the model to other sytems~\cite{Bormio}.
One observes however some discrepancies such as the behaviour of the high 
energy tail of the carbon energy spectrum already mentionned above and 
shown in Fig.~1b. This may be due to the sequential character of the 
process. The C fragments are generated at a later stage in the decay chain 
in both calculations. The kinetic energies they acquire are governed by
Coulomb barriers which are not high enough for these kinetic energies
to reach the values observed in the experimental tail of the distribution.

\bigskip
As already stated above, time is certainly correlated to the
characteristic features of the fragmentation process. It is, of
course, possible to estimate the duration of the decay. In Figs.~4 we
present the time evolution of the cumulated particle and fragments
yields. These figures show two interesting facts. The WTR rates lead
more quickly than the STR rates to the generation of light particles
and fragments, but after some time STR distributions start growing,
``catch up" with and even overshoot the WTR ones in the case of IMF
generation.
The STR scenario corresponds to a generalized fission
process which, at least at the beginning of the decay, is slower
than light particle evaporation. 

\bigskip
The second interesting and important fact concerns the duration
time of the process. An estimate of this time can be read from
Figs.~4 for both WTR and STR rates. When STR rates are used
the generation of light species Z=1 (resp. Z=3) is of the order of
$10^4$ {\small -} 2 $\times 10^4$ fm/c (resp. $10^4$ fm/c). The IMF 
generation takes also about $10^4$ fm/c. However, measured fragment 
correlations \cite{Lips} indicate that the characteristic correlation
time is approximately 80 fm/c or shorter. This strongly disagrees
with the STR results.

\bigskip
If the process is governed by WTR rates, the light species generation
is of the order of 5 $\times 10^3$ to $10^4$ fm/c for Z=1 and
5 $\times 10^2$ fm/c for Z=3. Most of the intermediate mass fragments
are already generated after a time interval which does not exceed 
300 fm/c. Hence the time over which IMF's are created is rather short, 
approximately three times larger than the experimentally estimated time.
But one should notice that the WTR rates may not be very realistic for
high excitation energies, see ref.~\cite{Richert2}.

\bigskip
The results obtained above show that STR rates generate a process which
is rather slow and for WTR rates the time scale is much shorter, although
both merge to similar final fragment size distributions as already 
mentionned above.

\bigskip
Whatever the degree of validity of the transition rates used in the present 
analysis, the shortness of the estimated experimental fragmentation time 
($\simeq 80$ fm/c) raises the question of the validity of the mechanism 
itself. Each decay of a given fragment into two new fragments is treated 
independently of the rest of the system in which this binary process occurs,
except for the conservation of total energy. This is conceptually difficult
to believe if the process is so quick and the description of the process
introduced by means of the WTR rates becomes doubtful. In this case
multiparticle correlations between fragments interacting through their 
mutual Coulomb interaction are certainly strong, and the barriers and decay
rates cannot be treated independently. Hence the quick decay must be 
governed by another mechanism.

\bigskip
In summary, the present binary decay model using WTR and STR decay rates 
reproduces well the experimental charge and energy distributions obtained 
for p + Au collisions at 8.1 GeV. As for the crucial observable, the 
time scale, it fails when one uses STR rates. In the case of WTR rates, the 
time scale is not too far from the experimentally estimated time but the 
validity of the mechanism itself is questionable.

\bigskip
\noindent
\large
{\bf Acknowledgments}\\
\normalsize

The authors would like to thank S.P. Avdeyev for his kind help and
interesting suggestions. They are grateful to D. Gross for a critical
reading of the text and constructive comments.

\newpage
\epsfig{file=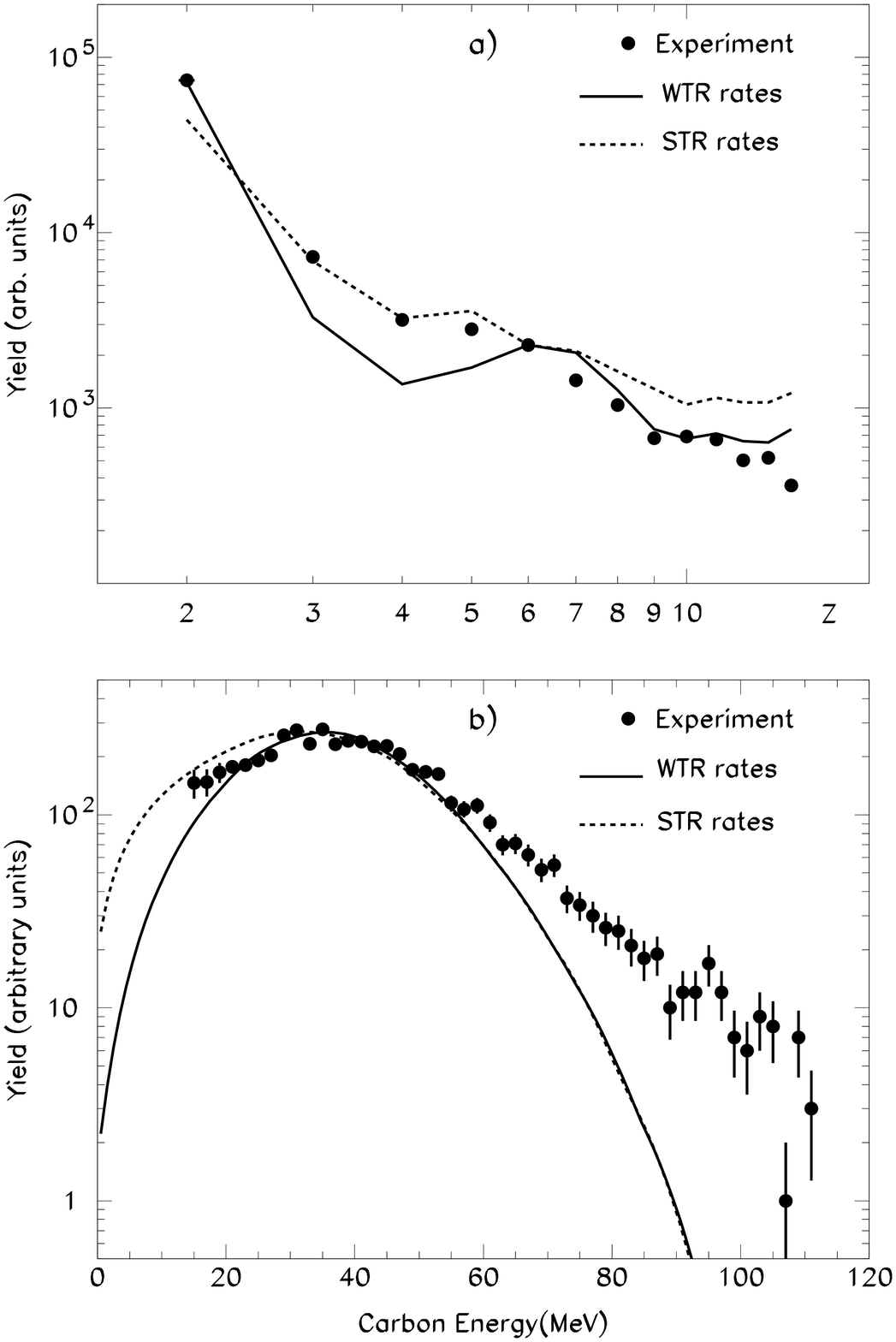,height=16cm}

\noindent
{\bf Fig.~1 :} (a) : Charge distribution of fragments; (b) : energy
distributions of carbon isotopes.

\newpage
\epsfig{file=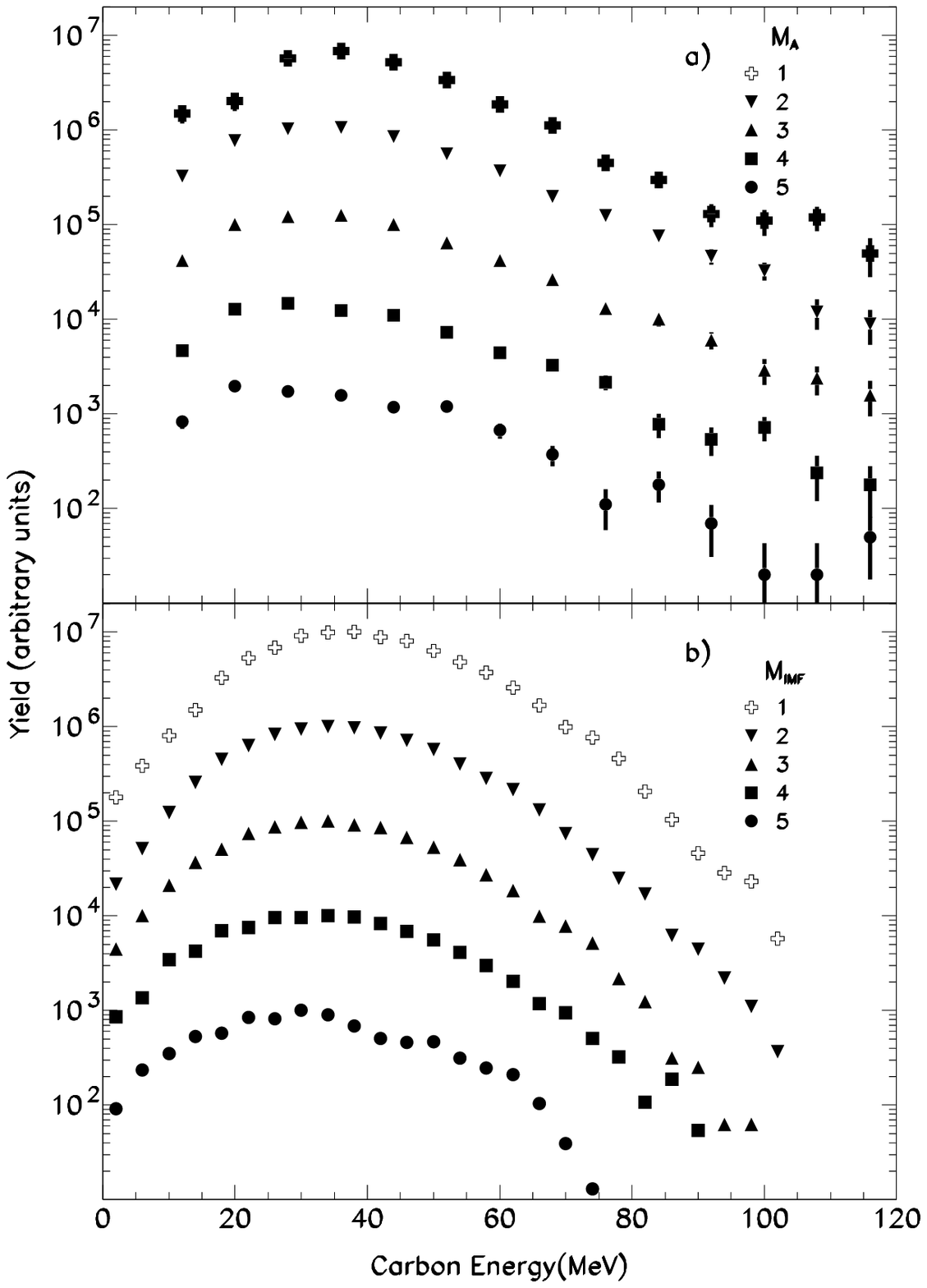,height=16cm}

\noindent
{\bf Fig.~2 :} Yield of carbon isotopes as a function of energy for different
intermediate mass fragments multiplicities (M$_{\mathrm{IMF}}$).
(a) : experiment ($M_{\mathrm{A}}$ is the number of measured IMF's);
(b) : calculation with WTR rates. Results obtained with STR rates are not
shown but very close.

\newpage
\noindent
\epsfig{file=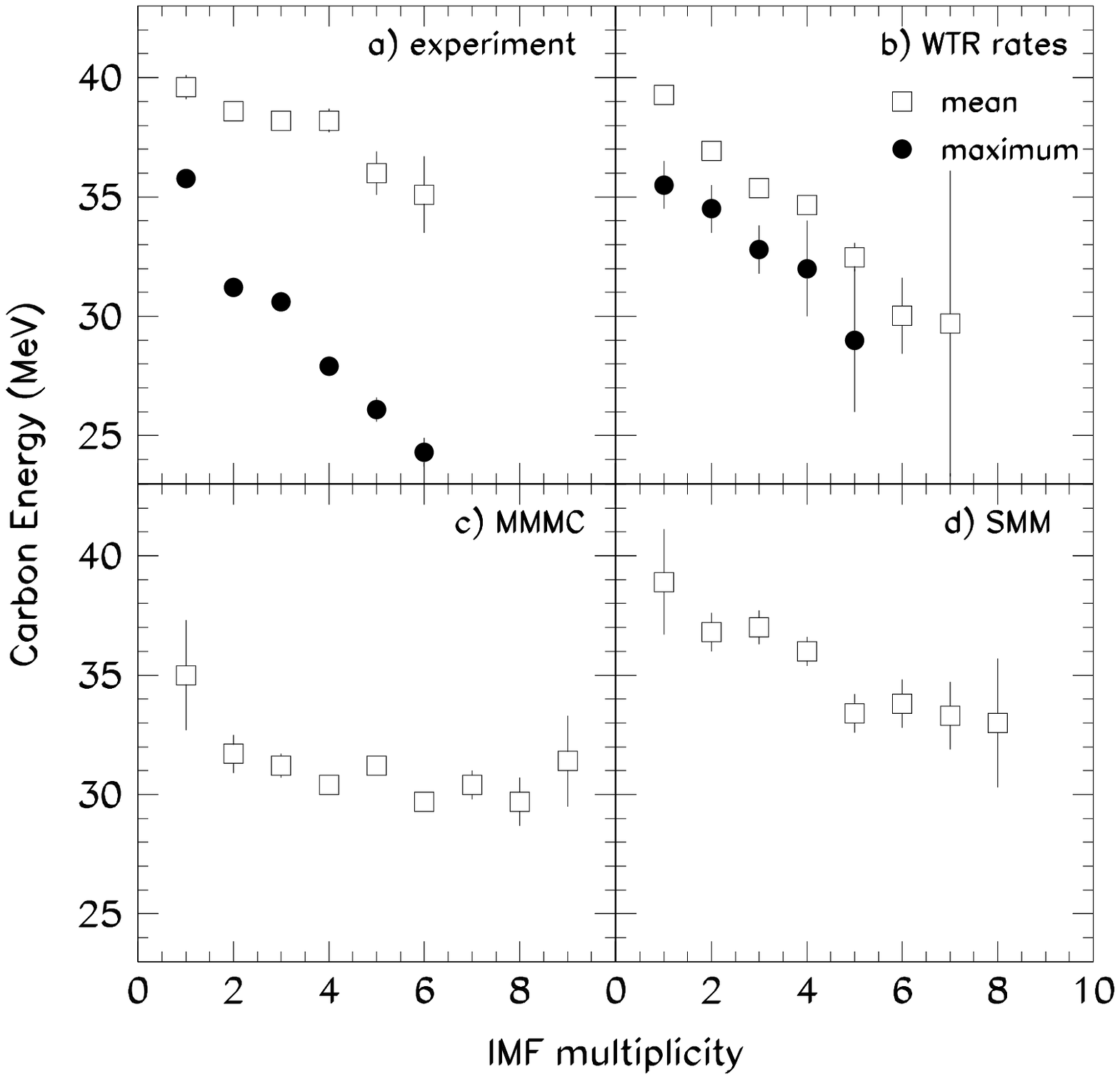,height=15cm}

\noindent
{\bf Fig.~3 :} Mean and maximum of the energy distribution of
carbon isotopes as a function of the multiplicity of intermediate
mass fragments (M$_{\mathrm{IMF}}$) : \\
(a) : Experiment; (b) : calculation with WTR rates for
excitation energies of 5 MeV/nucleon; (c) : mean
energy obtained with the MMMC model~\cite{Gross} for 5 MeV/nucleon;
(d) : mean energy obtained with the SMM model~\cite{Bondorf} for 5 MeV/nucleon.
E$_{\mathrm{mean}}$ is
defined as the average value obtained by integration over the
energy distribution of the carbons.

\newpage
\epsfig{file=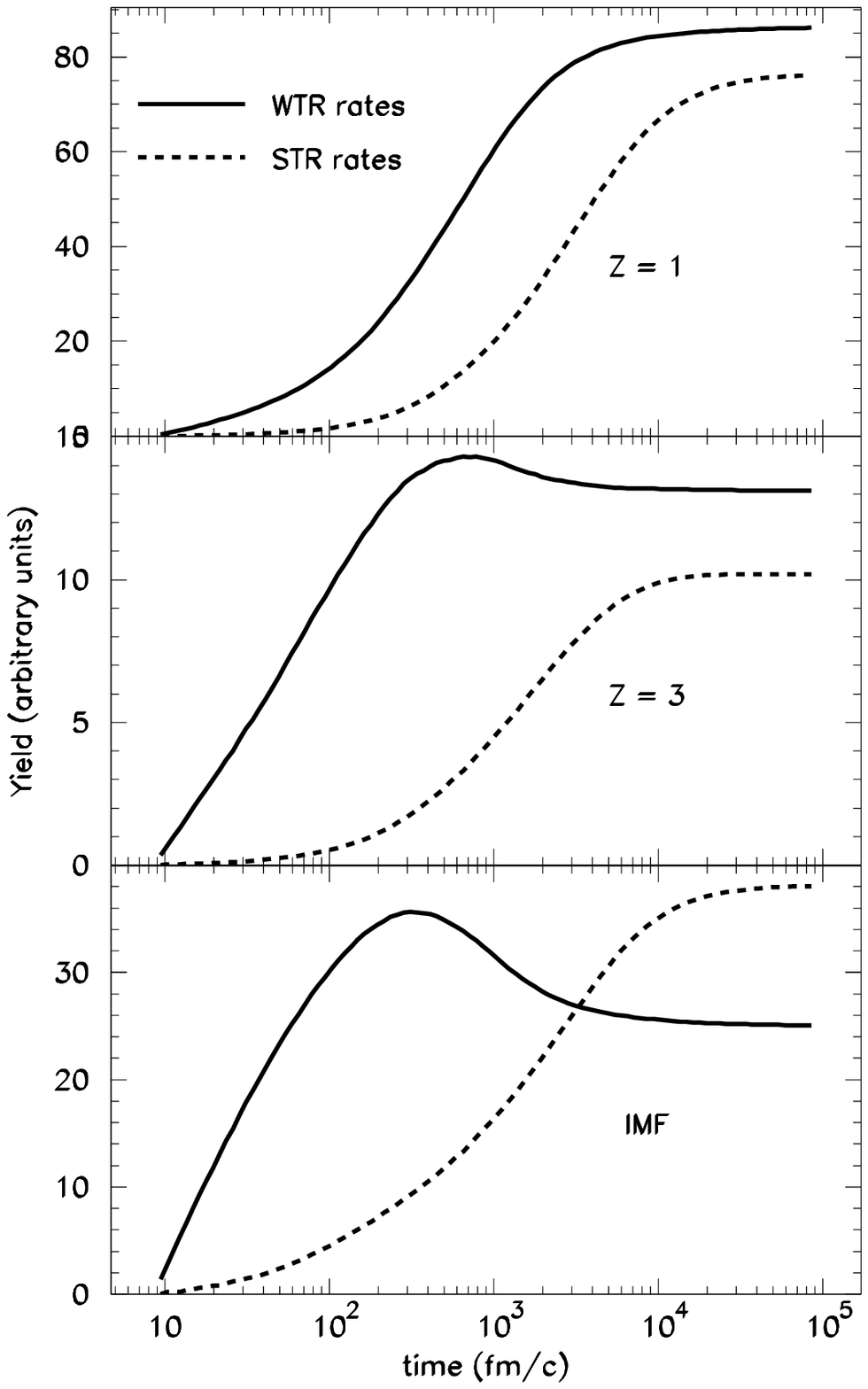,height=16cm}

\noindent
{\bf Fig.~4 :} Time dependence of the yields of particles and fragments
present in the system during the binary sequential decay process for
WTR and STR rates.

\end{document}